\documentclass[conference]{IEEEtran}
\IEEEoverridecommandlockouts
\usepackage{cite}
\usepackage{comment}
\usepackage{amsmath,amssymb,amsfonts}
\usepackage{algorithmic}
\usepackage{graphicx}
\usepackage{textcomp}
\usepackage{xcolor}
\usepackage{siunitx}

\usepackage{tikz}
\usetikzlibrary{arrows.meta,positioning,calc,decorations.pathreplacing}

\def\BibTeX{{\rm B\kern-.05em{\sc i\kern-.025em b}\kern-.08em
    T\kern-.1667em\lower.7ex\hbox{E}\kern-.125emX}}
\begin{document}

\title{ConamArray: A 32-Element Broadband MEMS \\ Ultrasound Transducer Array}
\author{
\IEEEauthorblockN{Dennis Laurijssen\IEEEauthorrefmark{1}\IEEEauthorrefmark{2}, 
Rens Baeyens\IEEEauthorrefmark{1}\IEEEauthorrefmark{2}, 
Walter Daems\IEEEauthorrefmark{1}\IEEEauthorrefmark{2}, 
Jan Steckel\IEEEauthorrefmark{1}\IEEEauthorrefmark{2}}
 \IEEEauthorblockA{\IEEEauthorrefmark{1}Cosys-Lab, Faculty of Applied Engineering, University of Antwerp, Antwerp, Belgium}
 \IEEEauthorblockA{\IEEEauthorrefmark{2}Flanders Make Strategic Research Centre, Lommel, Belgium\\
 \IEEEauthorrefmark{1}dennis.laurijssen@uantwerpen.be}
}

\maketitle

\begin{abstract}
This paper presents the ConamArray, a compact broadband ultrasound transducer array composed of 32 MEMS loudspeakers. Unlike conventional broadband transducers, which are typically large and require high driving voltages, the proposed array combines small form factor MEMS devices in a staggered two-row configuration to enable beam steering across a wide ultrasonic band. A dual-microcontroller back-end with synchronized multi-DAC outputs provides flexible waveform generation and runtime steering control.  

Both simulations and anechoic chamber measurements demonstrate that the ConamArray achieves stable beam steering, while also revealing the onset of grating lobes when steering to larger angles. These results confirm the feasibility of broadband beam steering using MEMS technology, opening new opportunities for applications in ultrasonic imaging, localization, and bio-inspired robotics.
\end{abstract}

\section{Introduction}
Airborne ultrasound transducer arrays are widely used in applications ranging from haptics, acoustic levitation~\cite{10.1371/journal.pone.0097590}, and manipulation, to biomimetic robotics~\cite{8794165, 10.1371/journal.pone.0054076, Steckel2015, doi:10.1073/pnas.1909890117} and sensing~\cite{10332222, 10214241, Haugwitz2024, Maier2021, 10210599, kerstens2019, 10833678, kerstens2019live, 8968469}. Central to many of these in-air ultrasound applications is the transducer's linearity, frequency range and its ability to control the spatial distribution of sound through beam steering and radiation pattern shaping. Conventional transducer technologies are, however, often limited in terms of bandwidth and scalability.

Broadband ultrasound transducers do exist, but their use in array configurations is constrained by several practical factors. Devices such as the Senscomp 7000 provide a extended frequency response, yet require high operating voltages and have a relatively large physical aperture. Due to their size relative to the wavelength, implementing beam steering with such transducers is highly impractical, as it inevitably introduces grating lobes. Other broadband technologies, such as ferroelectric film (e.g. Emfi~\cite{5653368}) transducers, exhibit a highly linear frequency response but similarly require high driving voltages and suffer from low acoustic emission amplitudes, which also limits their effectiveness in airborne ultrasound applications. Recent work from the University of Darmstadt~\cite{10784906, 10793522, 10793858} has demonstrated the use of additive 3D-printing methods to develop new ferroelectric transducers with improved emission levels. These devices show great promise as broadband ultrasound emitters for future research, yet they remain at an early stage of development.

In this work, we present a novel ultrasound device, ConamArray, that features a 32-element transducer array based on broadband MEMS speakers capable of beam steering. The system utilizes the newly developed USound UA-C0603-3T broadband MEMS speaker, which measures only \qty{6}{\milli\meter} in diameter and \qty{1.5}{\milli\meter} in height, while covering a wide frequency range from \qty{2}{\kilo\hertz} to \qty{80}{\kilo\hertz}. In contrast to conventional broadband devices, this compact MEMS-based array enables adjustable radiation patterns and a steerable main lobe without the need for high-voltage drive electronics.

The design, implementation, and characterization of ConamArray are described in the next sections of this paper. While the MEMS transducers are able to produce audible sounds, we wish to focus on their capabilities in the ultrasonic regime, highlighting their potential as a versatile (ultra)sound source for (bio-mimetic) robotics, acoustic experimentation, and other applications requiring broadband spatial control.

\section{Device overview}
The ConamArray consists of 32 USound UA-C0603-3T MEMS loudspeakers arranged in two staggered rows of 16 elements each. The physical dimensions of the MEMS transducers, together with the manufacturing constraints of the PCB, result in an inter-element spacing of \qty{6.1}{\milli\meter} within each row. Due to the staggered configuration, when the acoustic ports of both rows are projected onto the horizontal axis, the effective inter-element distance is halved. Based on this projected spacing, the onset frequency for grating lobes in the beam steering pattern was determined to be \qty{58.2}{\kilo\hertz}.

\begin{figure}
    \centering
    \includegraphics[width=1\linewidth]{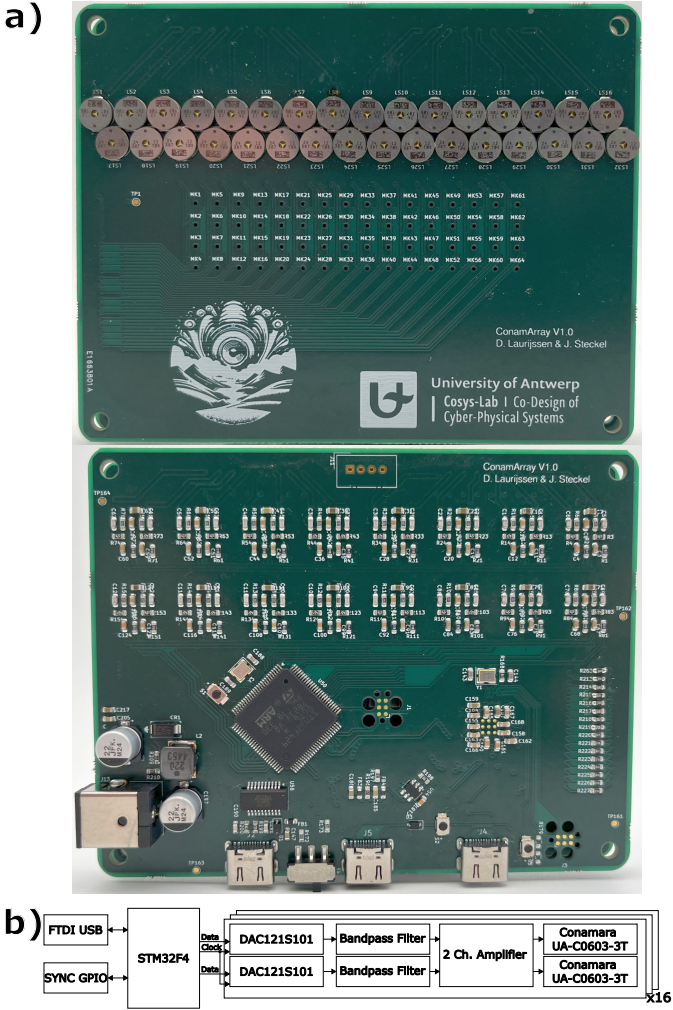}
    \caption{Both the front-end PCB featuring 32 UA-C0603-3T MEMS speakers can be seen in a) with the back-end PCB featuring the STM32F4 microcontroller, FTDI USB interface, \qty{32} TI DAC121S101 Digital-to-Analog Converters and \qty{16} operational amplifiers (ADA4099-2BRMZ) together with additional power circuitry is shown. The schematic representation of the interconnected front and back-end PCBs is shown in b).}
    \label{fig:hw_overview}
\end{figure}

The system is realized on two printed circuit boards (PCBs). The front-end PCB carries the transducer array, while the back-end PCB integrates the power regulation circuitry, a STMicroelectronics STM32F429 microcontroller, an FTDI FT231X Low-Speed USB interface, \qty{32} DAC121S101 converters, and \qty{16} ADA4099-2BRMZ operational amplifiers. Both boards are interconnected through 3 board-to-board connectors, forming a compact stacked assembly measuring \qty{102}{\milli\meter} by \qty{80}{\milli\meter} by \qty{9.5}{\milli\meter}. Photographs of the front-end and back-end PCBs are shown in figure~\ref{fig:hw_overview} (a), while figure~\ref{fig:hw_overview}(b) provides a hierarchical overview of the system architecture.

While the presented work focuses on its airborne ultrasound beam steering capabilities, the ConamArray was designed with pulse–echo functionality in mind. To this end, the front-end PCB also integrates a reception array consisting of 64 Knowles SPH0641LU4H-1 PDM MEMS microphones, arranged in a 16-by-4 regularly spaced layout. A Raspberry Pi RP2350B dual-core ARM-M33 microcontroller is used to control the synchronous acquisition of PDM microphone data and subsequent USB data transfers by utilizing the microcontroller's Programmable IO (PIO) state-machine blocks. High-speed streaming is enabled by an FTDI FT232H USB interface configured for synchronous FIFO operation. This architecture allows for real-time reception beamforming, extending the functionality of the system to pulse–echo applications.

To enable beam steering, the DACs are operated synchronously under a common sample clock generated by a timer peripheral of the STM32F429 microcontroller. On each rising edge of the clock, the microcontroller updates the parallel DAC inputs by means of DMA-controlled GPIO port operations, ensuring phase-coherent signal generation across all channels without CPU interaction. The output waveforms of the individual array elements are fully programmable at runtime. Through the FTDI USB interface, updated DAC output values can be streamed to the microcontroller, allowing the beam steering angle and radiation pattern to be dynamically adjusted. This design provides flexible control of the array’s acoustic output while maintaining precise synchronization among all elements.

\section{Methodology}
The beam steering capability of the ConamArray was investigated in three phases: numerical simulation, signal generation, and experimental validation.  

\subsection{Simulation}
\newcommand\dproj{d_{\text{proj}}}
\label{subsec:simulation}
We model the transmit array to (i) predict main-lobe steering, side-lobe behavior, and grating-lobe onset, and (ii) compute per-element \emph{time-domain delays} that are agnostic to the emitted waveform (chirp, multisine, sinusoidal burst, etc.). We adopt the right-handed coordinate system with $X$ the \emph{depth} (array normal, forward), $Y$ the \emph{horizontal} axis, and $Z$ the \emph{vertical} axis. The array lies approximately in the $Y$--$Z$ plane, i.e., $x_m \approx 0$ for all elements.

Let $c$ denote the speed of sound and $\mathbf{p}_m=[x_m,\,y_m,\,z_m]^{\mathsf T}$ the position of element $m$ with $m=0,\dots,31$ w.r.t. the array's center. We parameterize steering by elevation $\theta$ (angle from the $X$--$Y$ plane) and azimuth $\phi$ (angle of the projection of the steering vector in the $X$--$Y$ plane, w.r.t. the $X$ axis). The unit steering vector is
\begin{equation}
\mathbf{u}(\theta,\phi)\;=\;
\begin{bmatrix}
\cos\theta\cos\phi\\
\cos\theta\sin\phi\\
\sin\theta
\end{bmatrix},
\end{equation}
so that the \emph{transmit delay} for element $m$ (relative to the array center) becomes
\begin{equation}
\begin{aligned}
\tau_m
  &= -\frac{1}{c}\cdot\mathbf{p}_m^{\mathsf T}\cdot\mathbf{u}(\theta,\phi) \\
  &= -\frac{1}{c}\bigl(x_m\cos\theta\cos\phi
      + y_m\cos\theta\sin\phi
      + z_m\sin\theta\bigr).
\end{aligned}
\label{eq:tau_general}
\end{equation}

A special case follows directly which is called \emph{horizontal-plane steering} (no elevation), $\theta=0^\circ$ and $\phi$ arbitrary. This yields 
\begin{equation}
\tau_m = -\frac{m \cdot d_Y \cdot \sin\phi}{c},
\end{equation}
with $d_Y$ the interelement spacing along the $Y$ axis.

The two-row staggered geometry halves the effective projected pitch along $Y$ to $d_{\text{proj}} = d_Y/2$.
To avoid aliasing in the estimation of the delays in the array, we must ensure that:
\begin{align}
    \dproj \cdot\sin\phi < \frac{\lambda}{2}
\end{align}
This can be rewritten w.r.t. frequency as:
\begin{align}
    f < \frac{c}{2\cdot\dproj \cdot\sin\phi}
\end{align}
In the extreme azimuth case, i.e. $\phi = \qty{90}{\degree}$, this results in the broadside grating-lobe limit:
\begin{align}   
    \lambda &> 2 \cdot d_{\text{proj}} &\text{or}&&
    f &< \frac{c}{2\cdot\dproj}
\end{align} 
These relations agree with the simulated onset of grating lobes based on the measured geometry (projected spacing) and the reported value of $58.2\,$kHz.

With sampling rate $F_s$, each delay maps to a discrete sample delay
\begin{equation}
D_m = k_m = \lfloor\tau_m\cdot F_s\rfloor
\end{equation}
i.e. steering is implemented using integer shifts $k_m$. This time-domain approach is waveform-agnostic and simpler than frequency-domain phase steering for broadband calls, while maintaining inter-channel phase accuracy across the operating band. It must be noted that while this implementation was chosen, given that every transducer is driven by their dedicated DAC which offers great versatility in the output calls.

\subsection{Signal Generation}
The simulated emission signals are assembled into a 16-bit integer (zero padded 12-bit values) DAC input matrix $\mathbf{S}\in\mathbb{R}^{32\times N}$, where row $m$ contains the delayed waveform for element $m$. The STM32F429 updates the parallel inputs of all TI DAC121S101 converters synchronously using a timer-driven DMA routine. On each rising clock edge, all DAC data inputs are updated coherently, with a shared timer generated frame sync signal ensuring precise inter-channel phase control. Waveforms (and thus beam steering angles) can be reconfigured at runtime via the USB interface.  

\subsection{Experimental Setup}
Measurements were performed in an anechoic chamber. The ConamArray was mounted on a FLIR pan–tilt unit (PTU), while a calibrated microphone was placed at a distance of \qty{1.5}{\meter} on the PTU centerline, aligned with the midpoint between the two MEMS rows. This setup allowed direct comparison between simulated and measured beam profiles, quantifying steering accuracy, side-lobe levels, and the onset of grating lobes.  

\section{Results}
In this section we present both simulated and experimental results of the ConamArray. The focus is on the beam steering performance across the ultrasonic frequency range of \qty{20}{\kilo\hertz} to \qty{100}{\kilo\hertz}. Simulations were first conducted to predict the array response and to identify the onset of grating lobes for different steering angles. These predictions were then validated experimentally in the anechoic chamber. Both simulations and measurements are reported for azimuth steering angles of \ang{0}, \ang{-40}, and \ang{80}, enabling a direct comparison of the array’s performance under representative conditions.

\subsection{Simulated Beam Steering Performance}
To evaluate the beam steering capability of the two-row staggered geometry, we simulated the array response by varying the azimuth steering angle while sweeping frequency across the ultrasound band (\qty{20}{\kilo\hertz} to \qty{100}{\kilo\hertz}). In figure~\ref{fig:beam_patterns}, the plots in the top row show the simulated results where the horizontal axis represents the steering angle, the vertical axis represents frequency whereas the color scale indicates the radiated acoustic energy.  

At broadside ($\phi=\ang{0}$), the array produces a single dominant main lobe with minimal side-lobe energy across most of the spectrum with the projected element spacing ($d_{\dproj}=6.1/2=\qty{3.05}{\milli\meter}$). When steering the main lobe toward $\phi=\ang{-40}$, the beam broadens and the side-lobe level increases, with grating lobes becoming visible from around \qty{70}{\kilo\hertz}. At $\phi=\ang{80}$ steering, the main lobe becomes broader and the calculated onset grating lobe frequency of\qty{58.2}{\kilo\hertz} is strongly present, illustrating the strong degradation in beam quality when steering toward the edge of the front hemisphere. 

These results confirm that broadband beam steering is achievable, but the usable steering range is fundamentally limited by the onset of grating lobes and also illustrate the trade-off between steering range and grating-lobe suppression in a compact staggered array geometry. In practice, this implies that if grating lobes are to be avoided for operating in the ultrasonic regime we remain limited to modest steering angles, while the lower audible and ultrasonic frequencies support wider steering without significant artifacts. If further reduction of grating lobe effects is desired, a baffled approach in which acoustic waveguides are used to effectively reduce the element spacing has shown promise in the literature~\cite{jager2017air, rutsch2019wave, laurijssen2024broadband}. Such waveguides can be realized using CNC manufacturing techniques (e.g., 3D-printed acoustic channels) and could be incorporated in future iterations of the array design to extend the usable steering range.

\subsection{Experimental Results}
Experimental validation was performed in the anechoic chamber using the setup described in Section~\ref{subsec:simulation}. The ConamArray was mounted on a FLIR pan-tilt unit to align the array at defined azimuth angles, and a calibrated measurement microphone was placed at approximately \qty{1.5}{\meter} distance along the PTU centerline.  

\begin{figure}
    \centering
    \includegraphics[width=1\linewidth]{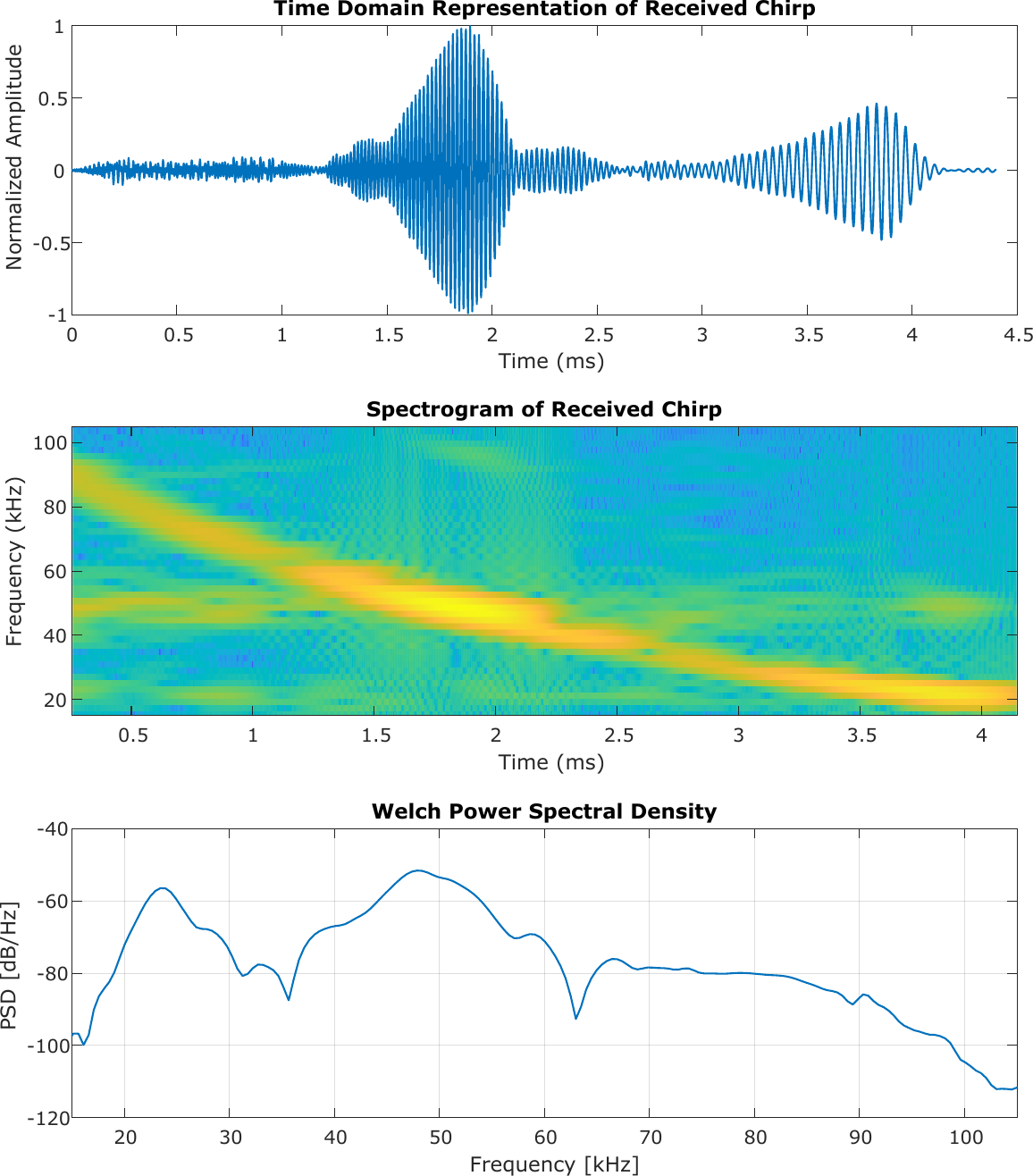}
    \caption{These plots show a recorded logarithmic chirp from \qty{100}{\kilo\hertz} to \qty{20}{\kilo\hertz} emitted from the ConamArray with its time-domain representation in a), the spectrogram representation in b) and the Welch Power Spectral Density in c).}
    \label{fig:freq_response}
\end{figure}

Figure~\ref{fig:freq_response} shows the measured frequency response of the array at broadside. The response confirms the broadband nature of the MEMS loudspeakers, with significant acoustic output spanning from \qty{100}{\kilo\hertz} to \qty{20}{\kilo\hertz}. The dip in the frequency spectrum near \qty{60}{\kilo\hertz} is an effect of destructive interference in the MEMS speaker cavity and could be alleviated somewhat by removing the protective film of the transducer with adverse ingress protection as a trade-off.

\begin{figure}
    \centering
    \includegraphics[width=1\linewidth]{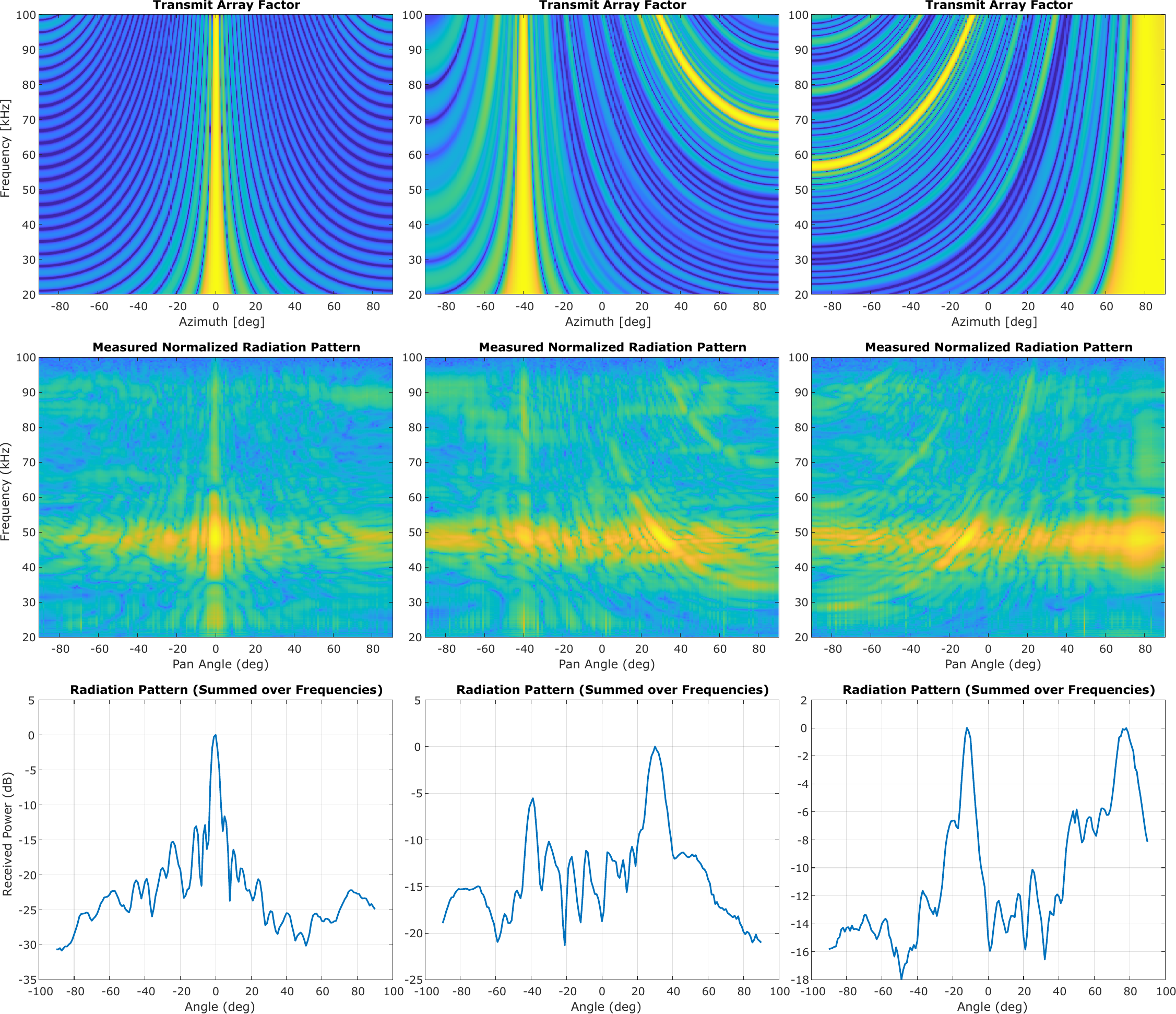}
    \caption{Results for simulated and measured data at steering angles $\phi=\ang{0}$, $\phi=\ang{-40}$, and $\phi=\ang{80}$. The top row shows simulated transmit array factors for a logarithmic chirp from \qty{100}{\kilo\hertz} to \qty{20}{\kilo\hertz}.  The middle row shows the normalized radiation pattern from the processed recorded data using the pan-tilt setup, and the bottom row shows the radiation pattern with received power summed over frequency.}
    \label{fig:beam_patterns}
\end{figure}

Measured beam patterns at steering angles of $\phi=\ang{0}$, $\phi=\ang{-40}$, and $\phi=\ang{80}$ are presented in figure~\ref{fig:beam_patterns}. At broadside ($\ang{0}$), the array produces a well-defined main lobe with minimal side-lobe energy consistent with simulation. When steering to $\ang{-40}$, the main lobe broadens and the side-lobe level increases, with grating lobes becoming visible around \qty{70}{\kilo\hertz} as expected. At $\ang{80}$ steering, the beam pattern's main lobe is strongly degraded, and grating lobes become more dominant in the ultrasonic band.  

These measurements closely match the simulated results, confirming that broadband beam steering is feasible but that the effective steering range in the ultrasonic regime is limited by the onset of grating lobes.

\section{Conclusion}
This work presented the ConamArray, a 32-element broadband MEMS ultrasound transducer array capable of dynamic beam steering across the ultrasonic frequency range of \qty{20}{\kilo\hertz} to \qty{100}{\kilo\hertz}. The array leverages compact USound MEMS loudspeakers in a staggered two-row configuration, combined with a synchronized multi-DAC driving architecture, enabling flexible waveform control and real-time steering.  

Simulations and anechoic chamber measurements demonstrated that the array achieves main-lobe formation and broadband steering. The onset of grating lobes was simulated and observed at approximately \qty{58.2}{\kilo\hertz} when steering to extremities of the frontal hemisphere, in agreement with the projected element spacing. Steering to moderate angles (e.g., \ang{-40}) resulted in a broadened main lobe and elevated side-lobe levels, while steering to extreme angles (e.g., \ang{80}) produced severe degradation of the beam pattern.  

These results confirm both the feasibility and the limitations of broadband beam steering with compact MEMS arrays. Future work will explore techniques for mitigating grating lobes, such as baffled designs using 3D-printed acoustic waveguides to reduce effective inter-element spacing, as well as extending the system to pulse-echo operation by exploiting the integrated MEMS microphone array. In addition, the current dual-microcontroller back-end will be reworked into a single FPGA-based control system, enabling centralized management of both transmission and reception paths with real-time in-the-loop processing and adaptive beam pattern adjustments. Such enhancements will broaden the applicability of the ConamArray in ultrasonic imaging, localization, and bio-inspired robotics experiments.

\clearpage

\bibliographystyle{IEEEtran}
\bibliography{citationsLibrary}

\begin{thebibliography}{10}
\providecommand{\url}[1]{#1}
\csname url@samestyle\endcsname
\providecommand{\newblock}{\relax}
\providecommand{\bibinfo}[2]{#2}
\providecommand{\BIBentrySTDinterwordspacing}{\spaceskip=0pt\relax}
\providecommand{\BIBentryALTinterwordstretchfactor}{4}
\providecommand{\BIBentryALTinterwordspacing}{\spaceskip=\fontdimen2\font plus
\BIBentryALTinterwordstretchfactor\fontdimen3\font minus \fontdimen4\font\relax}
\providecommand{\BIBforeignlanguage}[2]{{%
\expandafter\ifx\csname l@#1\endcsname\relax
\typeout{** WARNING: IEEEtran.bst: No hyphenation pattern has been}%
\typeout{** loaded for the language `#1'. Using the pattern for}%
\typeout{** the default language instead.}%
\else
\language=\csname l@#1\endcsname
\fi
#2}}
\providecommand{\BIBdecl}{\relax}
\BIBdecl

\bibitem{10.1371/journal.pone.0097590}
\BIBentryALTinterwordspacing
Y.~Ochiai, T.~Hoshi, and J.~Rekimoto, ``Three-dimensional mid-air acoustic manipulation by ultrasonic phased arrays,'' \emph{PLOS ONE}, vol.~9, no.~5, pp. 1--5, 05 2014. [Online]. Available: \url{https://doi.org/10.1371/journal.pone.0097590}
\BIBentrySTDinterwordspacing

\bibitem{8794165}
D.~Laurijssen, R.~Kerstens, G.~Schouten, W.~Daems, and J.~Steckel, ``A flexible low-cost biologically inspired sonar sensor platform for robotic applications,'' in \emph{2019 International Conference on Robotics and Automation (ICRA)}, 2019, pp. 9617--9623.

\bibitem{10.1371/journal.pone.0054076}
\BIBentryALTinterwordspacing
J.~Steckel and H.~Peremans, ``Batslam: Simultaneous localization and mapping using biomimetic sonar,'' \emph{PLOS ONE}, vol.~8, no.~1, pp. 1--11, 01 2013. [Online]. Available: \url{https://doi.org/10.1371/journal.pone.0054076}
\BIBentrySTDinterwordspacing

\bibitem{Steckel2015}
------, ``Spatial sampling strategy for a 3d sonar sensor supporting batslam,'' \emph{IEEE International Conference on Intelligent Robots and Systems}, vol. 2015-December, pp. 723--728, 12 2015.

\bibitem{doi:10.1073/pnas.1909890117}
\BIBentryALTinterwordspacing
R.~Simon, S.~Rupitsch, M.~Baumann, H.~Wu, H.~Peremans, and J.~Steckel, ``Bioinspired sonar reflectors as guiding beacons for autonomous navigation,'' \emph{Proceedings of the National Academy of Sciences}, vol. 117, no.~3, pp. 1367--1374, 2020. [Online]. Available: \url{https://www.pnas.org/doi/abs/10.1073/pnas.1909890117}
\BIBentrySTDinterwordspacing

\bibitem{10332222}
A.~García-Requejo, M.~Pérez-Rubio, A.~Hernández, W.~Wright, and L.~Marnane, ``Ultrasonic device-free localisation system modelling for performance analysis,'' in \emph{2023 13th International Conference on Indoor Positioning and Indoor Navigation (IPIN)}, 2023, pp. 1--7.

\bibitem{10214241}
E.~Aparicio-Esteve, J.~Ure{\~n}a, {\'A}.~Hern{\'a}ndez, and J.~M. Villadangos, ``Combined infrared-ultrasonic positioning system to improve the data availability,'' \emph{IEEE Sensors Journal}, vol.~23, no.~20, pp. 25\,152--25\,164, 2023.

\bibitem{Haugwitz2024}
C.~Haugwitz, F.~Krauß, G.~Allevato, M.~Rutsch, J.~H. Dörsam, S.~Wismath, S.~Soennecken, A.~Harth, C.~M. Heyl, C.~Othmani, S.~Merchel, M.~E. Altinsoy, T.~Hahn-Jose, and M.~Kupnik, ``A 220 khz air-coupled spiral ultrasonic phased array using waveguides,'' \emph{IEEE Ultrasonics, Ferroelectrics, and Frequency Control Joint Symposium, UFFC-JS 2024 - Proceedings}, 2024.

\bibitem{Maier2021}
T.~Maier, G.~Allevato, M.~Rutsch, and M.~Kupnik, ``Single microcontroller air-coupled waveguided ultrasonic sonar system,'' \emph{Proceedings of IEEE Sensors}, vol. 2021-October, 2021.

\bibitem{10210599}
G.~Allevato, C.~Haugwitz, M.~Rutsch, R.~Müller, M.~Pesavento, and M.~Kupnik, ``Two-scale sparse spiral array design for 3d ultrasound imaging in air,'' \emph{IEEE Open Journal of Ultrasonics, Ferroelectrics, and Frequency Control}, vol.~3, pp. 113--127, 2023.

\bibitem{kerstens2019}
R.~Kerstens, D.~Laurijssen, and J.~Steckel, ``ertis: A fully embedded real time 3d imaging sonar sensor for robotic applications,'' in \emph{2019 International Conference on Robotics and Automation (ICRA)}, 2019, pp. 1438--1443.

\bibitem{10833678}
J.~Steckel, P.~R. Parra, A.~Aerts, D.~Laurijssen, W.~Jansen, W.~Daems, and J.~Barber, ``Fl-rtis, a novel multimodal sensor using high-speed camera and active 3-d sonar for insect ensonification,'' \emph{IEEE Sensors Letters}, vol.~9, no.~2, pp. 1--4, 2025.

\bibitem{kerstens2019live}
R.~Kerstens, G.~Schouten, W.~Jansen, D.~Laurijssen, and J.~Steckel, ``Live demonstration of ertis, an embedded real-time imaging sonar sensor,'' in \emph{2019 IEEE SENSORS}.\hskip 1em plus 0.5em minus 0.4em\relax IEEE, 2019, pp. 1--1.

\bibitem{8968469}
R.~Kerstens, D.~Laurijssen, G.~Schouten, and J.~Steckel, ``3d point cloud data acquisition using a synchronized in-air imaging sonar sensor network,'' in \emph{2019 IEEE/RSJ International Conference on Intelligent Robots and Systems (IROS)}, 2019, pp. 5855--5861.

\bibitem{5653368}
S.~Epure, R.~Belea, and D.~Aiordachioaie, ``Emfi based ultrasound transceivers,'' in \emph{2010 IEEE 16th International Symposium for Design and Technology in Electronic Packaging (SIITME)}, 2010, pp. 117--122.

\bibitem{10784906}
A.~A. Altmann, S.~Suppelt, O.~Ben~Dali, B.~Latsch, D.~Spiehl, S.~Zhukov, F.~Herbst, J.~H. Dörsam, A.~Blaeser, and M.~Kupnik, ``Single-step 3d printing of flexible ferroelectret sensors with large air cavities,'' in \emph{2024 IEEE SENSORS}, 2024, pp. 1--4.

\bibitem{10793522}
S.~Suppelt, A.~A. Altmann, S.~Schaumann, N.~Demuth, M.~Müller, L.~E. Jazdzewski, T.~E. Gómez~Álvarez Arenas, C.~Bretthauer, A.~Bittner, and M.~Kupnik, ``Acoustical sensitivity and linearity of an air-coupled 3d-printed ferroelectret ultrasonic receiver,'' in \emph{2024 IEEE Ultrasonics, Ferroelectrics, and Frequency Control Joint Symposium (UFFC-JS)}, 2024, pp. 1--4.

\bibitem{10793858}
A.~A. Altmann, S.~Suppelt, M.~Ruhl, S.~Schaumann, B.~Latsch, O.~Ben~Dali, S.~Zhukov, D.~Flachs, X.~Zhang, C.~Thielemann, H.~Von~Seggern, and M.~Kupnik, ``Monolithic wideband air-coupled ultrasonic transducer based on additively manufactured ferroelectrets,'' in \emph{2024 IEEE Ultrasonics, Ferroelectrics, and Frequency Control Joint Symposium (UFFC-JS)}, 2024, pp. 1--4.

\bibitem{jager2017air}
A.~J{\"a}ger, D.~Gro{\ss}kurth, M.~Rutsch, A.~Unger, R.~Golinske, H.~Wang, S.~Dixon, K.~Hofmann, and M.~Kupnik, ``Air-coupled 40-khz ultrasonic 2d-phased array based on a 3d-printed waveguide structure,'' in \emph{2017 IEEE International Ultrasonics Symposium (IUS)}.\hskip 1em plus 0.5em minus 0.4em\relax IEEE, 2017, pp. 1--4.

\bibitem{rutsch2019wave}
M.~Rutsch, A.~J{\"a}ger, A.~Unger, T.~Kaindl, and M.~Kupnik, ``Wave propagation in an acoustic waveguide for ultrasonic phased arrays,'' in \emph{2019 IEEE International Ultrasonics Symposium (IUS)}.\hskip 1em plus 0.5em minus 0.4em\relax IEEE, 2019, pp. 796--799.

\bibitem{laurijssen2024broadband}
D.~Laurijssen, W.~Daems, and J.~Steckel, ``Broadband mems microphone arrays with reduced aperture through 3d-printed waveguides,'' in \emph{2024 IEEE SENSORS}.\hskip 1em plus 0.5em minus 0.4em\relax IEEE, 2024, pp. 1--4.

\end{thebibliography}

\end{document}